\documentclass[useAMS,usenatbib]{mn2e}
\usepackage{graphicx}

\usepackage{subfigure}

\title[Low Angular Momentum Outflows]{Hierarchical formation of bulgeless galaxies:  \\ Why outflows have low angular momentum.}
\author[C. B. Brook et al.]{C. B. Brook$^{1}$,  F. Governato$^{2}$ , R. Ro\v{s}kar$^{2,3}$, G. Stinson$^{1}$, A. Brooks$^{4}$ , J. Wadsley$^{5}$,   
\newauthor  T. Quinn$^{2}$, B. K. Gibson$^{1}$, O. Snaith$^{1}$, K. Pilkington$^1$ , E. House$^1$, A. Pontzen$^{6}$\\ 
        $^1$University of Central Lancashire, Jeremiah Horrocks 
            Institute for Astrophyics \& Supercomputing,
            Preston, PR1~2HE, UK \\
        $^2$Astronomy Department, University of Washington, 
            Box 351580, Seattle, WA 98195-1580, USA \\
	$^3$  Institute for Theoretical Physics, University of Z\"{u}rich, Winterthurerstrasse 190, Z\"{u}rich, Switzerland\\
          $^4$  California Institute of Technology, M/C 130-33, Pasadena, CA 91125, USA\\
                  $^5$Department of Physics and Astronomy, McMaster University, Hamilton, ON L8S 4M1, Canada\\
                  $^6$Institute of Astronomy and Kavli Institute for Cosmology Cambridge, Madingley
Road, Cambridge CB3 0HA, UK\\}

\begin{document}

\date{}

\pagerange{\pageref{firstpage}--\pageref{lastpage}} \pubyear{2002}

\maketitle

\label{firstpage}

\begin{abstract}

  Using high resolution, fully cosmological smoothed particle
  hydro-dynamical simulations of dwarf galaxies in a $\Lambda$ cold
  dark matter ( $\Lambda$CDM) Universe, we show how high redshift gas outflows can modify the baryon angular momentum  distribution and allow pure disc galaxies to form.  We outline how
  galactic outflows preferentially remove low angular momentum
 baryons, due a combination of a) star formation peaking at high
  redshift in shallow dark matter potentials, an epoch when accreted gas has relatively low angular momentum,   b) the existence of an extended
  reservoir of high angular momentum gas in the outer disc to provide
  material for prolonged SF at later times and c) the tendency for
  outflows to follow the path of least resistance which is
  perpendicular to the disc. We also show that outflows are enhanced
  during mergers, thus expelling much of the gas which has lost its
  angular momentum during these events, and preventing the formation
  of ``classical'', merger driven bulges in low mass, field
  galaxies. Stars formed prior to such mergers form a low surface
  brightness, extended stellar halo component, similar to those
  detected in nearby dwarfs.
 \end{abstract}


\section{Introduction}

The angular momentum of disc galaxies is thought to originate from
tidal torques imparted by surrounding structures in the expanding
Universe \citep{peebles69,barnes87}, prior to proto-galactic
collapse. Disc galaxies will form if gas gains a similar amount of
angular momentum as the dark matter in this process, and if this
angular momentum is subsequently retained as  the gas cools to the
centres of dark matter halos and settles into a disc
\citep{fall80}. Gas subsequently fragments and forms stars.  Within
cold dark matter (CDM) cosmology this picture of the origin of
galactic angular momentum is believed to hold even as galaxies are
built hierarchically, with the added assumption that mergers result in
building bulge components as discs are destroyed by violent relaxation
\citep{barneshernquist96}.

However, detailed modelling has highlighted problems with this
scenario. In simulations, gas cools to the centre of the earliest
collapsing structures making them very centrally concentrated.
Dynamical friction occurring during the mergers of such systems
results in significant loss of angular momentum
\citep{navarrosteinmetz00,maller02}. This problem has largely been
overcome in simulations by increasingly effective recipes for feedback
from supernovae \citep{thacker00,stinson06,g07}, and increased
resolution that decrease spurious angular momentum losses
\citep{mayer08,sales10}. Despite this encouraging progress, decreasing
the impact of dynamic friction does not solve all the problems of
galaxy formation which relate to angular momentum within hierarchical
structure formation.  In addition, the angular momentum distribution
of baryons in real galaxies differs significantly from the
distribution within dark matter halos \citep{vdBS01,dutton09}.  Even
if angular momentum were fully conserved, CDM halos have a low angular
momentum tail \citep{bullock01}.  Recent simulations show that
galaxies formed in a CDM Universe have realistic disc sizes
\citep{brooks10}, however the close coupling of dark matter and gas
while acquiring their angular momentum implies that baryons should
also have a substantial low angular momentum component, which is
generally believed to be reflected in the bulge components
\citep{vdbosch01,stinson10}.  Yet many real galaxies have no bulge;
$\sim 70\%$ of galaxies with stellar mass of $<10^9$M$_\odot$ have a
Sersic index of less than 1.5 \citep{dutton09b} and  many more massive
galaxies do not have classical bulges \citep{kormendy10,fisher10}. Pure disc structures reflect an underlying angular momentum distribution which is seemingly at odds
with theories of galaxy formation within the accepted $\Lambda$CDM
paradigm \citep{bullock01,vdbosch01}.  Furthermore, in a hierarchical
CDM Universe, mergers drive gas toward the centre of DM halos, making
the formation of low angular momentum bulge stars an apparently
inevitable consequence.  The persistence and degree of these problems
means that the manner in which disc galaxies attain their angular
momentum distribution has been  considered one of the major long standing
challenges to the $\Lambda$CDM paradigm.

The removal of low angular momentum gas in galactic outflows is one
viable solution to explain the discrepancy in angular momentum
distributions between dark matter and baryons \citep{dekel86,binney01}
and the main focus of this work. Several lines of evidence suggest
that supernova driven galactic winds are able to expel large amounts
of gas from galaxies during the galaxy formation process
\citep{matthews71,veilleux05}. Galactic winds have been observed, and
extensively studied in local star forming galaxies
(e.g. \citealt{Lynds63,Axon78,martin99,Ohyama02,Martin05,Rupke05}), or
inferred from the metallicity distribution of stellar populations in
low mass galaxies \citep{kirby11}. Outflows are expected to be more
common at high redshift where star formation is more active
\citep{madau96}. Indeed, outflows have been detected at high redshift
from Lyman-break selected galaxies
\citep{pettini98,Simcoe02,Shapely03,adelberger05}, as well as
gravitationally lensed galaxies \citep{pettini2000}.

We recently simulated dwarf galaxies which match a wide range of
observed dwarf galaxy properties, including the absence of a bulge and
cored DM distribution that causes a linearly rising rotation curve
with no inner ``peak'' (\citealt{governato10}; G10 hereafter). The
simulated galaxies lie simultaneously on the Tully-Fisher relation,
and the size-luminosity relation \citep{brooks10}.  Using techniques
which closely mimic observations, the galaxies have been shown to have
mass distributions and stellar masses which are consistent with
galaxies in the THINGS survey (\cite{oh10}, Figure 5) of similar
rotational velocity and fall on the baryonic Tully-Fisher relation, as
derived by \cite{mcgaugh05}.  Most importantly, the simulated dwarfs
also match the distribution of angular momentum of stars observed in
pure disc galaxies, differing significantly from that of the dark
matter halo in which they are embedded (see Figure~4 of G10, as well
as Figure~\ref{vdB} of this paper).  These properties indicate that
our simulations can provide unique insights into the acquisition and
retention of angular momentum of disc galaxies.  Here we provide an
overview of the various interlinked processes which determine the
angular momentum distribution of stars in our simulated bulgeless disc
galaxy. In particular we outline why the significant amount of gas
which is expelled from the galaxy is primarily low angular momentum
material.

In Section~\ref{code}, we review the code and the simulation initial
conditions. We present the properties of our simulated galaxy in
Section~\ref{galaxies}, highlighting resolution issues as well as the
effects of allowing gas to cool to dense regions before forming
stars. We plot star formation rates and outflow rates, as well as
their ratio (the ``mass loading'') in Section~\ref{sfr_out}. We show
that the outflows have a strong bias toward low angular momentum in
Section~\ref{bias}. We outline the reasons for outflows having
preferentially low angular momentum material: early accreted material
has low angular momentum, yet needs to escape a relatively low
potential well (Section~\ref{early}); the existence of an extended
reservoir of high angular momentum gas (Section~\ref{extended}); with
outflows being perpendicular to the disc (Section~\ref{perp}).  We
also show that the starburst triggered in the late merger expels the
bulk of the gas which lost its angular momentum during this event
(Section~\ref{bulge}). Our summary and discussion follows in  Section~\ref{summary}. \\

\section{The Simulation.}\label{code}
At z $=$ 0 the virial mass of the halo in the dark matter only run of
our simulated galaxy is 3.7 $\times$ 10$^{10}$M$_{\odot}$ (the virial
mass is measured within the virial radius R$_{\rm vir}$, the radius
enclosing an overdensity of 100 times the cosmological critical
density). The halo was selected within a large scale, low resolution,
simulation run in a concordance, flat, $\Lambda$-dominated cosmology:
$\Omega_0=0.24$, $\Lambda$=0.76, $h=0.73$, $\sigma_8=0.77$, and
$\Omega_{b}=0.042$ \citep{spergel03, verde03}.  We evolved the
simulation using the fully parallel, N-body+smoothed particle
hydro-dynamics (SPH) code GASOLINE which self consistently models gas
hydrodynamics and cooling, star formation, stellar mass loss, energy
feedback from supernovae, metal enrichment and gas diffusion together
with structure formation within theCDM model.  A detailed description
of the code and of the simulations is available in the literature
\citep{wadsley04,governato10,shen10}.  The version of the code used in
this study includes radiative cooling and accounts for the effect of a
uniform background radiation field on the ionization and excitation
state of the gas. The cosmic ultraviolet background is implemented
using the Haardt-Madau model \citep{haardtmadau96}, we use a standard
cooling function for a primordial mixture of atomic hydrogen and
helium at high gas temperatures, but we include the effect of metal
cooling for T $<$ 10$^4$ \citep{bromm01,mashchenko08}. The possible
effects of including H$_2$ metal line cooling and SF linked to H$_2$
abundance and some preliminary results are briefly discussed in \S5.

Star formation occurs when cold gas reaches a given threshold density
\citep{stinson06}, which is dependent on the volume of the star
forming regions that can be resolved
\citep{saitoh08,robertson08,tasker08}. We used 100 amu/cm$^3$ in the
main run analysed here. The adoption of a higher density threshold for
star formation (most previous studies used 0.1amu/cm$^3$) has the
effect of limiting SF to gas clumps similar in mass to real star forming
regions and of increasing drastically the amount of energy per unit
mass released into the gas directly affected by supernova
feedback. This higher threshold is justified by the high mass and
spatial resolution of this run that allow individual star forming
regions to be resolved. Star formation then proceeds at a rate
proportional to $\rho_{gas}^{1.5}$, i.e. locally enforcing a Schmidt
law. The adopted feedback scheme is implemented by releasing thermal
energy from supernovae into the gas surrounding each star particle
\citep{stinson06}. The energy release rate is tied to the time of
formation of each particle (which effectively ages as a single stellar
population with a Kroupa IMF).  To model the effect of feedback at
unresolved scales, the affected gas has its cooling shut off for a
time scale proportional to the Sedov solution of the blastwave
equation, which is set by the local density and temperature of the gas
and the amount of energy involved. In the main run described in
this study this translates into regions of $\sim$ 150 to 400 pc in
radius being heated by feedback from multiple, overlapping supernovae,
and having their cooling shut off for typically 5-10 million years.
However, even during high z starbursts only a few percent of the gas
in the disc plane is heated by supernova feedback to temperatures $>$
40,000 K.  The effect of feedback is to regulate star formation
efficiency as a function of mass \citep{brooks07}. Only two other
parameters are needed, the star formation efficiency ($\epsilon$SF =
0.1) and the fraction of supernova energy coupled to the inter stellar
medium (ISM) (eSN = 0.4).  Similar values have been used in previous
works by this group \citep{g07}. However, here we slightly increased
$\epsilon$SF from 0.05 to 0.1 to ensure a better normalization to the
local Schmidt law. We explored values of eSN as high as 0.6 and
cooling shutoff times changing by a factor of a few to verify that
results are robust to small changes in the description of SF.

\begin{figure}%
\hspace{-1.2cm} \includegraphics[height=.26\textheight]{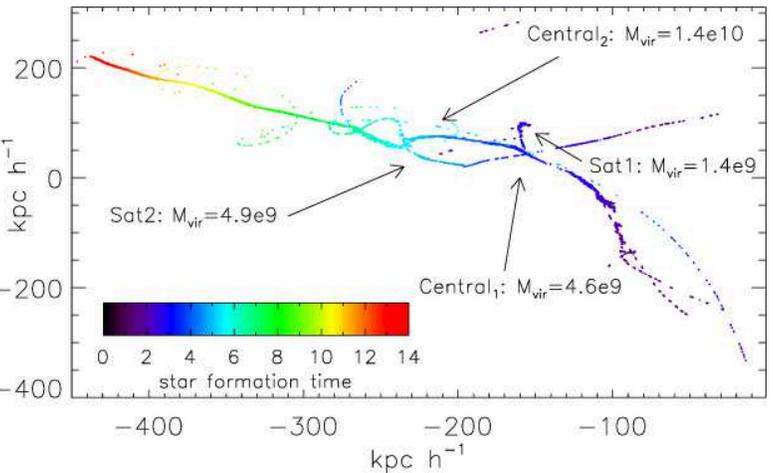}
  \caption{A type of merger tree. We plot the birth position within the simulation box of every star which ends up in the final galaxy at z=0. We are tracing the birth of stars as the galaxy moves through the simulation box. The axis are the x and y co-ordinates of the simulation box, shown in co-moving kpc. The colors are the birth time of the stars, where $z=0$ corresponds to 13.7 Gyrs.. We see in this plot the merger events which involve satellites which are large enough to host stars.  Two significant merger events in particular can be seen at $\sim 2.5$ Gyrs and $\sim7$ Gyrs.  For these events, we note the virial mass of the satellites (Sat1 and Sat2) prior to the merger, as well as the mass of the main progenitor at corresponding times. The passages of Sat2 around the central galaxy and  its extended merging period is also evident.}
\label{merger}
\end{figure}

\subsection{The Galaxy: Resolution and Cosmic Variance}\label{galaxies}
The simulation presented here is part of an ongoing campaign to study
the formation and evolution of galaxies in a $\Lambda$CDM
cosmology. Recent work has highlighted the role of numerical
resolution in driving the structural properties of simulated galaxies
and specifically their internal mass distribution
\citep{mayer08,governato10}.  We analysed the simulated galaxy used in
this study at different resolutions and showed that our key results
are retained at high resolution regimes, which have spatial resolution
less than $\sim200$ pc and gas particle mass less than a few
$10^4$M$_\odot$ (G10). Yet we emphasise the importance of resolving
star forming regions by allowing gas to collapse to high density
regions during these merger events. For example a high SF threshold
DG1MR forms about 3$\times$10$^8$ M$_\odot$ in stars.  This relatively low star
formation efficiency leads to an M$_{stars}$/M$_{halo}$ ratio of
$\sim$ 1/200, which is close to what is measured for nearby small
galaxies with resolved kinematics and photometry \citep{oh10}.  On
the other hand a 0.1 amu/cm$^3$ threshold allows eight times more
stars to form (see G10, online material).  Adopting high resolution
significantly decreases SF also in halos with mass $>$ 10$^{11}$
M$_\odot$ (Governato et al, in prep, Mayer et al in prep), reducing the
overproduction of stars \citep{brooks10} and bringing them close to
estimates for more massive galaxies \citep{guo10} that predict
M$_{stars}$/M$_{halo}$ $\sim$ 0.04-0.05. The importance of resolving star forming regions was also demonstrated by the results from a constrained merger simulation done by
\cite{teyssier10}. The more localised star formation
within an inhomogeneous inter stellar medium is crucial in our model
in terms of allowing large scale outflows to be generated (G10).

\begin{figure}%
\hspace{-.5cm}  \includegraphics[height=.23\textheight]{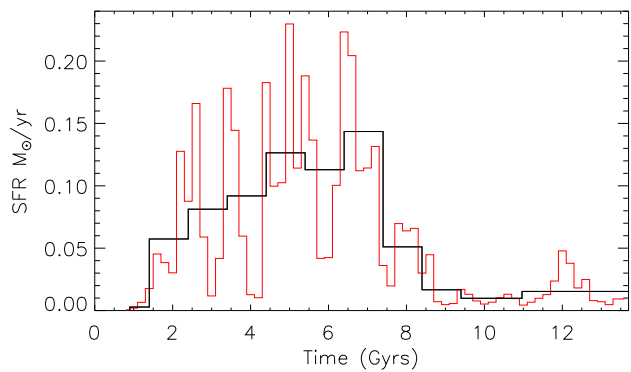}
  \caption{The star formation rate plotted as a function of time. We show 200 Myr bins in red, highlighting the bursty nature of the star formation. The black line shows 1 Gyr bins, which is shown because that matches the time resolution information that we have for outflows, which are less well defined than star formation. The peak in star formation at $\sim7$Gyrs is associated with the last major merger.   }
\label{starformation}
\end{figure}

\begin{figure}%
\hspace{-.5cm} \includegraphics[height=.23\textheight]{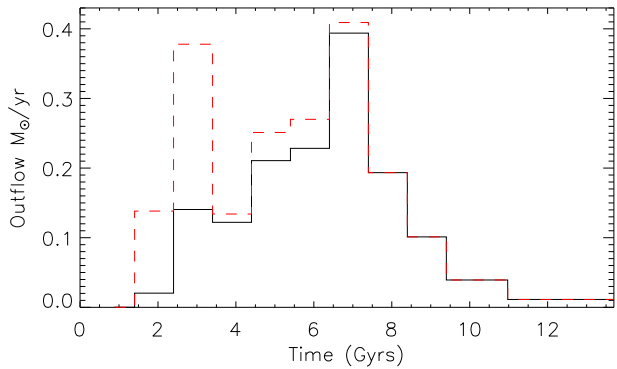}
  \caption{The outflow rate of gas from the galaxy as a function of time. The black line shows the outflows from the central galaxy. The red dashed line adds outflows from the two most significant satellites accreted during the galaxy's evolution. The peak outflows occur during the merger events at $\sim 2.5$ Gyrs and $\sim7$ Gyrs.   }
\label{outflows}
\end{figure}

\begin{figure}%
\hspace{-.5cm}  \includegraphics[height=.23\textheight]{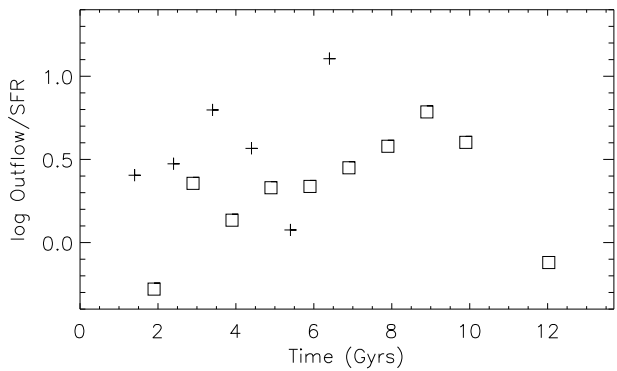}
  \caption{The evolution of the ratio of the mass of expelled gas to the mass of stars  formed, showing  a `mass loading' from the central galaxy (square symbols) of greater than one for almost the entire simulation, and a rise associated with a late major merger event. The crosses show the mass loading of the major satellite which merges at $z\sim 1$, which also shows significant outflows.    }
\label{massloading}
\end{figure}

\begin{figure}%
\hspace{-.5cm} \includegraphics[height=.23\textheight]{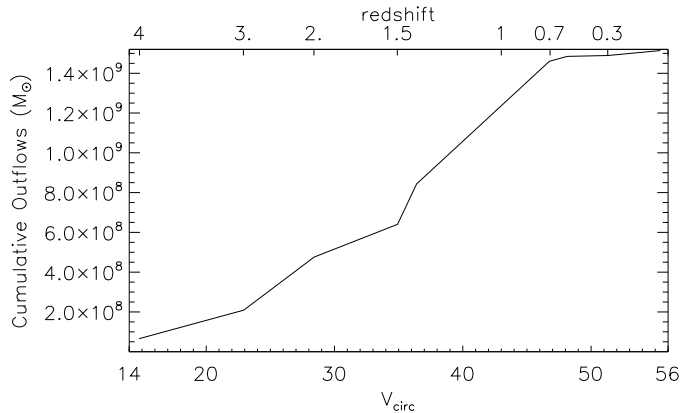}
  \caption{The cumulative mass of outflows as a function of circular velocity. Significant outflows occur prior to the simulated galaxy attaining all its material. The top axis indicates the corresponding redshifts.}
\label{cumulative}
\end{figure}

From the simulation we generated frequent outputs in order to
determine outflow rates and the properties of the gas prior to being
blown out. To this end, we use DG1MR from G10, which has outputs every
$\sim 500$ Myrs, and has an effective resolution $2304^3$ particles in
a 25$\;$Mpc box, with gravitation softening of 110pc and initial
stellar mass of $\sim 1700$M$_\odot$. The morphology (that of a
bulgeless dwarf irregular galaxy), star formation rate, final stellar
and baryon fraction, M$_{HI}$/L$_B$, rotation velocity, magnitude and
color show little change between DG1 and DG1MR. These properties
contrast significantly compared to the same simulation run with a low
threshold for star formation (DG1LT), which does not generate large
scale outflows.  Details of the properties of the galaxies DG1, DG1MR
and DG1LT using techniques which mimic observations were presented in
G10.  DG1MR has M$_i=-16.9$, $g-r=0.54$,
M$_{star}=5.3\times10^{8}$M$_{\odot}$, scalelength R$_s=0.9$kpc,
rotational velocity V$_{rot}=55$kms$^{-1}$, and ratio of neutral hydrogen to B-band luminosity M$_{HI}$/L$_B=1.0$.  \\

Bulges are commonly asserted to be formed in mergers, with galaxies
which have more mergers, or mergers with larger mass ratios, assumed
to have larger bulges (e.g. \citealt{barneshernquist96,hopkins10}). In CDM
cosmologies, merger histories of low mass galaxies are equally as
diverse and rich as high mass galaxies, yet it is only in low mass
field galaxies where bulgeless discs are common \citep{dutton09b}
(dwarf satellites often have spheroidal morphologies e.g.
\citealt{mayer01}). To validate the CDM model, simulations should form
low mass field galaxies with low B/D ratios and rising rotation curves
starting from most realizations of cosmological initial conditions. It
is important to highlight that the success of our analysed galaxy is
not a result of a particular, special merging history. We showed in
G10 the results of two simulations which had very different merging
histories, highlighting that we have not selected a particularly
quiescent simulated galaxy. In Figure~\ref{merger}, we plot a type of
merger tree for the simulation analysed in this study, which shows the
birth position within the simulation box of every star which ends up
in the final galaxy at z=0. We are tracing the birth of stars as the
galaxy falls through the simulation box. The axes are the x and y
co-ordinates of the simulation box, shown in co-moving kpc. The colors
are the birth time of stars. We see in this plot the merger events
which involve satellites which are large enough to host stars.  Two
significant merger events in particular can be seen at $\sim 2.5$ Gyrs
and $\sim7$ Gyrs. Prior to merging, Sat1 has a virial mass of
$1.4e9$M$_{\odot}$, gas mass of $2.0e8$M$_{\odot}$ and stellar mass of
$3.7e6$M$_{\odot}$, with the central galaxy having virial mass of
$4.6e9$M$_{\odot}$, gas mass of $2.6e8$M$_{\odot}$ and stellar mass of
$4.1e7$M$_{\odot}$ at a corresponding time.  Also evident are the
passages of Sat2 around the central galaxy and hence its merging
period which extends for $\sim 1.6$ Gyrs between $\sim 5.2$Grys
$(z=1.2)$ and final coalescence at $\sim 6.8$ Gyrs $(z=0.8)$.  Prior
to merging, Sat2 has a virial mass of $4.9e9$M$_{\odot}$, gas mass of
$3.4e8$M$_{\odot}$ and stellar mass of $3.8e7$M$_{\odot}$, while the
central galaxy has virial mass of $1.4e10$M$_{\odot}$, gas mass of
$1.3e9$M$_{\odot}$ and stellar mass of $1.8e8$M$_{\odot}$ at a
corresponding time.

\section{Star Formation and Outflows}\label{sfr_out}

In Figure~\ref{starformation} we plot the star
formation rate as a function of time for our simulated galaxy. In red,
we show the star formation using 200 Myr bins. The bursty nature of
star formation is apparent. The black line shows 1 Gyr bins, to allow
a better match to the time resolution we have for our outflows in
subsequent plots. Star formation shows bursts, some of which are
associated with mergers, with a high star formation rate at $\sim 2.5$
Gyrs associated with Sat1, and then peaks again at $\sim 7$ Gyrs. This
final increase in the star formation rate is associated with the last
major merger (Sat2), which has a mass ratio of $\sim 3:1$. Subsequent
to the end of this merger, the star formation rate decreases
significantly, and is quite low for the final $\sim 4$ Gyrs but
continues to be bursty.

\begin{figure}%
\hspace{-.5cm} \includegraphics[height=.23\textheight]{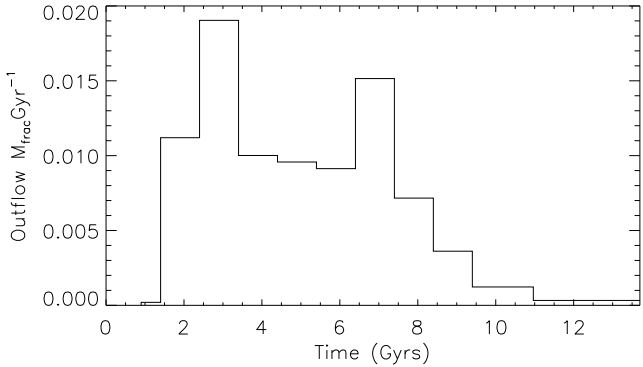}
  \caption{The evolution of the ratio of the mass of  expelled gas to the total virial mass of the galaxy. The ratio peaks at early times, with the subsequent steady decline interrupted by a rise which occurs during the last major merger event.}
\label{outermass}
\end{figure}

The outflow rate, in M$_{\odot}\:yr^{-1}$, is shown in
Figure~\ref{outflows}. The black line shows the outflows from the
central galaxy. The red dashed line adds outflows from the two most
significant satellites accreted during the galaxy's evolution, marked
Sat1 and Sat2 in Figure~\ref{merger}. Outflows are calculated in a
very simple manner: any gas which is bound and within $5R_*$ at any
output timestep, but is outside $R_{vir}$ of the final galaxy at
$z=0$, is assumed to have been expelled, where $R_*$ is the radial
distance to the furthest young star (age $< 200$ Myrs) from the centre of the
galaxy. Typically, $5R_*$ is $\sim 0.5 R_{vir}$. Our results are not
sensitive to this choice, so long as the radius is well beyond the
star forming region which means that we do not miss gas which is being
expelled, and is not too close to $R_{vir}$, which can result in gas
which is not bound being falsely identified as outflows.  We only
include gas which does not return to the galaxy at later times in our
definition of outflows in this study. Using 28 outputs which are
$\sim500\;$ Myrs apart, each outflow particle is traced through every
output to find the maximum ``jump'' in radial distance from the
centre. The jump in radius is invariably associated with a temperature
that shows that the gas particle was heated by supernova energy. The
time of the radial and temperature jump determines the outflow time
for the particle.

We note that the degree and direction of outflows in our simulation is
not directly imposed, as in other feedback schemes (e.g
\citealt{springel03,oppenheimer06}), but occur
naturally due to thermal pressure exerted on gas in our formulation of
supernova feedback.

The general shape of the outflows histogram is similar to the star
formation rate, including a particularly prominent peak in outflows
during the coalescence of the last major merger event at time $\sim$
$7\;$ Gyrs $( z\sim0.8)$. This correspondence between a merger induced
starburst and subsequent outflows is discussed in Section~\ref{bulge}.

The ratio of the outflows from the central galaxy to the star
formation rate is called the ``mass loading'' or the ``loading
factor''. The log of the ratio of the outflows from the central galaxy
to the star formation rate is shown in Figure~\ref{massloading}
(squares). The mean mass loading from the central galaxy throughout
the simulated galaxy's evolution is 2.3, with a significant increase
associated with the merger at $\sim7\;$ Gyrs. We will see that a high
mass loading is a crucial ingredient that determines the final angular
momentum content of our simulated galaxy. In Figure~\ref{massloading}
we also show the mass loading of the outflows from the major
satellite, Sat2 (crosses), which merges to the central galaxy at $z
\sim 1$. This highlights that outflows also occur from the satellites
which are massive enough to from stars.  From here on, we analyse
properties of outflows from the central galaxy, or main progenitor, at
each time and simply refer to ``outflows''.

In Figure~\ref{cumulative} we plot the cumulative mass of such
identified outflows as a function of circular velocity at the time of
expulsion. It is apparent that significant amounts of gas is
expelled while the galaxy is still relatively low mass, but that the
galaxy continues to accrete material and grow. To highlight this, we
plot the evolution of the ratio of outflows to the total mass of the
galaxy (M$_{frac}$) in Figure~\ref{outermass}. This plot shows a peak
in the fraction of mass being ejected at early times, between $2-3\;$
Gyrs, when the potential of the proto galaxy is relatively low and the
outflow rate is high. A subsequent steady decline in M$_{frac}$ is
halted only during the last merger at $\sim 7\;$ Gyrs, during which
large outflows occur.

\begin{figure}%
\hspace{-.5cm} \includegraphics[height=.24\textheight]{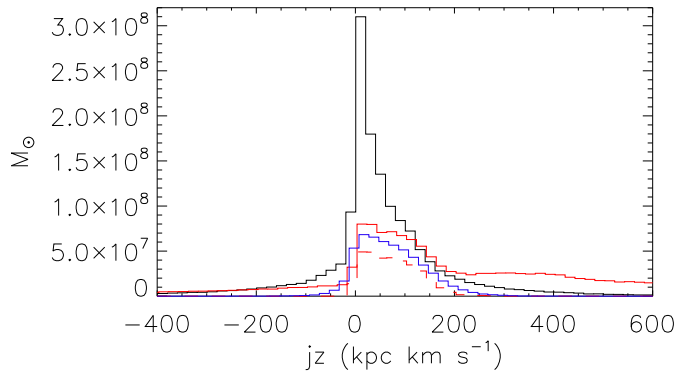}
\caption{The angular momentum distributions of stars (blue), all gas (red) and HI gas (red dashed) within the virial radius at $z=0$  and gas which is blown out of the galaxy throughout its evolution (black). A bias for low angular momentum to be expelled is evident, as is the fact that a large amount of gas is expelled.}%
\label{jzhist}%
\end{figure}

\section{Selective removal of  low angular momentum in outflows}\label{bias}

As a consequence of the strong outflows 70\% of the baryons are expelled by winds during the galaxy's evolution,
and there is a significant bias toward low angular momentum in the
expelled gas. Figure~\ref{jzhist} compares the $j_z$ distribution of
gas which has been ejected in outflows (black line) to the $j_z$
distribution of all baryons within the virial radius of the galaxy at
$z=0$, showing gas (red), HI gas (red dashed), and stars (blue). Here,
$j_z$ is the component of the angular momentum whose vector is
perpendicular to the disc.  The angular momentum of the outflowing
gas is determined at the timestep prior to its heating and
ejection from the central galaxy.  The shape of the angular momentum
distribution of the ejected material differs strongly from the
material which makes up the galaxy at $z=0$. The outflow comprises a
large low angular momentum peak. The outflows also display a
significant tail of negative angular momentum material. By contrast,
the warm/hot gas in the galaxy at $z=0$ contains high angular momentum
material, while the cold (HI) gas and stars have very little negative
angular momentum material and relatively little low angular momentum
material.  We note that a very similar result occurs when total
angular momentum is plotted, which we omit for brevity.

 \begin{figure}%
\hspace{-.5cm} \includegraphics[height=.23\textheight]{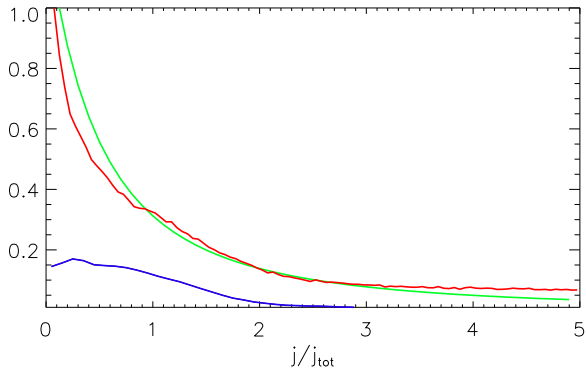}
  \caption{The final distribution of $j = rVc(r)$ for ``observable'' baryons in the simulation, which consists of  stars and HI gas (blue). In red we plot the distribution of all baryons which would have ended in the galaxy if outflows were not present, i.e. it is the distribution of stars, HI gas and  warm/hot gas within the virial radius  at $z=0$, combined with outflows  whose angular momentum is determined at the timestep prior to being heated and expelled.  An analytic fit, as determined in Bullock et al. (2001),   to the typical distribution for dark matter halos is overplotted (green).}
\label{vdB}
\end{figure}

\begin{figure}%
\hspace{-1cm} \includegraphics[height=.24\textheight]{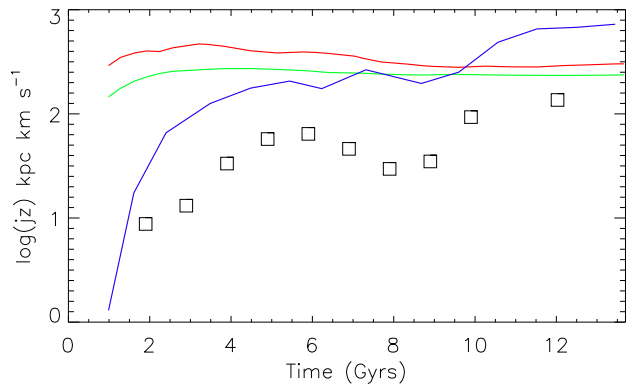}
\caption{The evolution of the mean  angular momentum ($j_z$)  of baryons (red) and dark matter (green) which are within the virial radius at $z=0$, compared with the mean angular momentum of gas which is being ejected during the galaxy's evolution (squares). Also shown is the mean angular momentum of gas which is being accreted at each time (blue line).  Material which collapses early into the galaxy has relatively low angular momentum, and is available for being expelled, and this  is replaced by inflowing gas which has progressively higher angular momentum.}%
\label{jztime}%
\end{figure}

In order to make a more direct link to observations, Figure~\ref{vdB}
shows the angular momentum distribution, $j/j_{tot}$, for
``observable'' baryons, defined as a combination of stars and cold
(HI) gas in order to mimic \cite{vdBS01}, who plot such a distribution
for several observed galaxies. Here, $j_{tot}$ is the mean angular
momentum of these ``observable'' baryons, and the angular momentum is
derived using $j = rV_{c}(r)$, where $r$ is the radius and $V_c$ the
circular velocity.  We also plot the distribution of all baryons as
they would be in the absence of outflows (red line). That is to say,
the distribution of outflows, as measured just prior to the time of
ejection, plus all baryons within the virial radius at $z=0$.  This
distribution is also normalised by the same $j_{tot}$ of the disc
baryons, in order to allow direct comparison with the blue line which
represents the distribution of observable baryons. A very different
distribution is apparent between what is observed at $z=0$, and the
total baryonic component in absence of outflows. In particular a large
amount of low angular momentum gas is present due to the expelled low
angular momentum gas, as well as a high angular momentum tail of
warm/hot gas.

Also shown on this plot is the distribution of angular momentum,
normalised by the mean angular momentum, for a typical dark matter
halo, as characterised in \cite{bullock01} and used in
\cite{vdBS01}. The total baryons which would have been in the galaxy
in the absence of outflows (red line) have a very similar shape to
this distribution, although we note that the normalization, $j_{tot}$,
that we used was the mean angular momentum of the disc
baryons. Normalising by the mean of the total angular momentum of
baryons which would have been in the galaxy at $z=0$ would shift the
distribution to the left.  Nevertheless, the shape of the distribution
of all baryons which would have been in the galaxy in the absence of
outflows, and that of dark matter halos, are remarkably similar,
highlighting the role that outflows play in causing the distribution
of observed baryons to differ so sharply from their host halos.

We now identify the processes which result in the strong bias toward
low angular momentum of outflows that we have identified.  In
combination they result in the ejection of low angular momentum
material and a significant alteration of the angular distribution of
baryons, and in particular the distribution of stars and cold (HI) gas
within galaxies compared to their host dark matter halos.

\subsection{Early accreted material has low angular momentum.}\label{early}
Material which falls into the central regions of proto-galaxies at the
earliest times has been subjected to the least amount of torquing from
the large scale structure \citep{ryden88,quinn92}. Material falling in
at later times has higher angular momentum.  This results in early
outflows having less angular momentum than later outflows, and
introduces a bias whereby low angular momentum material is preferentially expelled.
Figure~\ref{jztime} shows the evolution of $j_z$ of outflows
(squares), calculated at the time of ejection, as well as the inflows
(blue line) at a corresponding time. Also shown is the evolution of
$j_z$ for the baryons (red line) and dark matter (green line) which
end up in the galaxy at $z=0$. A similar plot results when total
angular momentum is plotted in this manner. The central galaxy, or
main progenitor, is used as a reference frame at all times, and as we
stated above, outflows are those ejected from the central galaxy,
while inflows, described below, are also to the central galaxy.

After spinning up, the dark matter material which ends in the galaxy
at $z=0$ essentially retains its angular momentum. The baryons which
end up in the galaxy at $z=0$ have also retained their angular
momentum.  At the early times, it is low angular momentum material
that has been accreted and is hence available for blow out and this is
reflected in the very low angular momentum of material which is
ejected at early times. Only gas which has been accreted can be
ejected, so a large bias is clear between ejected gas and the gas
which will end up in the galaxy at $z=0$.  The low star formation
rates and relatively high blowout rates (i.e. the high mass loading,
see Figure~\ref{massloading}) means that a large fraction of low
angular momentum material is not forming stars. As outer, high angular
momentum shells of gas are accreted, they make up a larger fraction of
the angular momentum distribution of gas than they would if none of
the earlier accreted gas were expelled. {\it This naturally alters the
  distribution of gas angular momentum compared with dark matter. Low
  star formation efficiency and high gas fractions mean that blown out
  gas has preferentially low angular momentum}.

The blue line traces the angular momentum of inflowing gas at each
time. We define inflowing material as gas which has reached $5R_*$ for
the first time, in order to make a fair comparison with the outflowing
material as we have defined it (recall that $R_*$ is the radial
distance to the furthest young star from the centre of the galaxy at each
time). Clearly, inflowing gas has significantly higher angular
momentum than the outflows at all times. At late times, the inflowing
material has higher angular momentum than the mean of all gas which is
within the virial radius at $z=0$, reflecting the fact that later
accreted material has high angular momentum.

Lower potential wells at early times favour higher blowout rates, at a
time when lower angular momentum material is being accreted.

\subsection{Extended reservoir of high angular momentum material.}\label{extended}

\begin{figure}
  \includegraphics[height=.5\textheight]{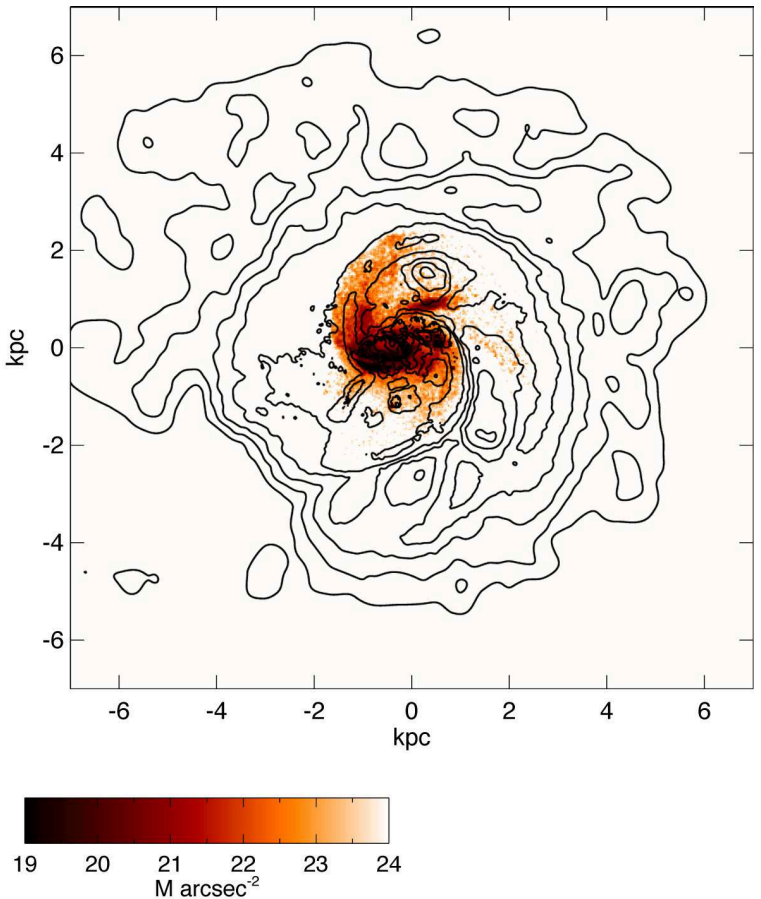}
  \caption{The face-on B-band surface brightness map of the galaxy at $z\sim1.2$, over plotted by an HI contour map. The HI contours range from $10^{19}cm^{-2}$ to $10^{22}cm^{-2}$ in steps of $10^{0.5}cm^{-2}$.
  The extent of the star forming region is reflected in the B-band map, while the extended nature of the HI is evident.   }
  \label{reservoir}
\end{figure}

\begin{figure*}%
\begin{center}
 \includegraphics[height=.27\textheight]{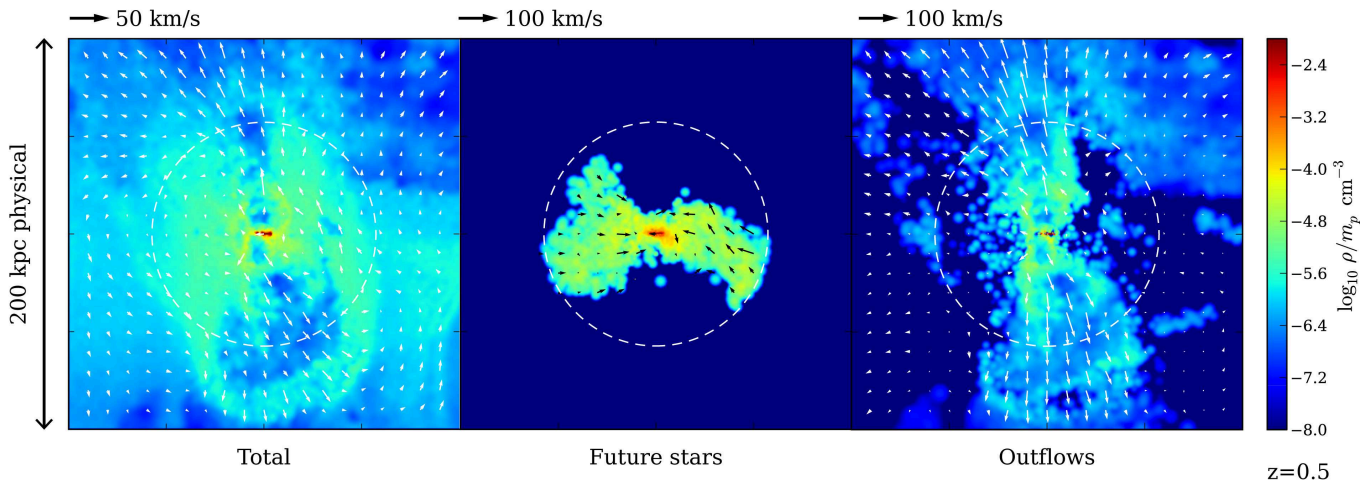}
 \caption{Left Panel: The background is a density map of all gas in
   the simulated galaxy at $z=0.5$, where the disc is oriented edge
   on. We have taken a 200x200 kpc box. The dashed line marks the
   virial radius. The arrows are velocity vectors, indicating the
   direction of outflows, with the size of the arrows related
   magnitudes of velocities as indicated above the panel.  Middle
   Panel: The gas which is feeding star formation, i.e. we include
   only gas at $z=0.5$ which will subsequently form stars by $z=0$. We
   use the same scale and again indicate the virial radius. The gas
   feeding the star formation is accreted largely from low angles from
   the plane of the disc. Right Panel: Here we plot outflowing gas,
   again with velocity vectors demonstrating that outflows are
   strongly directed perpendicular to the disc.}
\label{flow}
\end{center}
\end{figure*}

Star formation only occurs in the inner regions of the galaxy where
the gas reaches sufficiently high densities. Extended HI gas discs are
found around local isolated low mass galaxies \citep{broeils97}. Such
low mass, isolated galaxies have gas dominating their baryonic mass
fractions, with values typically above $70\%$, and as high as $95\%$
\citep{schombert01,geha06} There is evidence that extended HI discs
occur in intermediate \citep{puech10} and high redshift galaxies
\citep{daddi10}. Further, there is evidence for the existence of
warm/hot gas in the halos of disc galaxies
\citep{spitzer56,spitzer75,sembach03}.  In our simulations, gas beyond
the star formation region serves as a repository of high angular
momentum material.  Typically, between 30\% and 40\% of gas within the
virial radius of our simulated galaxy lies within the star forming
region at a given time.  Figure~\ref{reservoir} shows the face-on
B-band surface brightness map of the galaxy at $z\sim1.2$, over
plotted by an HI contour map.  The extent of the star forming region,
from which outflows are driven, is reflected in the B-band map, while
the extended nature of the HI is evident. At this time, 47\% of all
gas is beyond the star forming region. Gas within the star forming
region has average specific angular momentum of $jz_{mean}=66$ kpc
kms$^{-1}$ while gas beyond the star forming region has
$jz_{mean}=131$ kpc kms$^{-1}$.  Looking at just the HI gas at this
time, 20\% of HI is beyond the star forming region, and has
$jz_{mean}=145$ kpc kms$^{-1}$. Gas which gains energy directly from
supernovae thus has relatively low angular momentum, so simply
modelling this gas which is within the star forming regions as being
blown out will preferentially remove low angular momentum gas.

\subsection{Outflows perpendicular to the disc.}\label{perp}
Our simulations, as well as observations of outflows, indicate that as
gas is expelled, it can also entrain gas which is in outer regions of
the galaxy, and sweep it out with the outflows
\citep{stewart2000,schwartz04,veilleux05}. The relative importance of
these two modes of outflow are not well constrained. Direct expulsion
from inner regions appears to be the dominant mode
\citep{hawthorn2007}, and is also the main mechanism of gas removal in
our simulations. Here we show that that, consistent with observations
\citep{heckman87,bland88,shopbell98,veilleux02,hawthorn2007} and other
theoretical models \citep{maclow99,pieri07}, the outflows from the
central regions are preferentially perpendicular to the plane of the
disc.

Figure~\ref{flow} shows the direction of gas flows from our galaxy at
$z\sim 0.5$, when a minor starburst occurs. The color gas density map
shows the edge on view of the galaxy, with velocity vectors
overplotted showing the direction and magnitude of the gas
velocity. The virial radius is marked by a dashed white line. The right
hand panel shows the outflows, and reveals that they follow
the path of least resistance perpendicular to the disc.  These
outflows thus preferentially entrain material in regions which are perpendicular to
the disc.  The highest angular momentum material, which is in the
extended gas disc surrounding the star forming regions, is the least
affected by the outflows. We measure a low value for the mean angular
momentum in the plane of the disc of this material which is identified
as outflows, with $jz_{mean}(outflows)=20$kpc kms$^{-1}$, while total
angular momentum is high, $j_{mean}(outflows)=193$kpc kms$^{-1}$, as
the material is getting driven to large radii, perpendicular to the
plane, at high velocities.  Here, we have used $j_{mean}\equiv
\sqrt{jx_{mean}^2+jy_{mean}^2+jz_{mean}^2}$.

By contrast, gas which feeds star formation  falls on the
galaxy from a direction which is in the plane of the disc, and away from the outflows. The central
panel of Figure~\ref{flow} shows the edge on galaxy, again at $z=0.5$,
but this time only gas which will form stars by $z=0$ is plotted,
again with vectors indicating the velocity of the gas.  The gas which
feeds star formation is clearly shown to flow primarily from the plane
of the disc.  This gas has a relatively high planar angular momentum,
$jz_{mean}(star feed)=79$kpc kms$^{-1}$, which is a large fraction of
this material's total angular momentum $j_{mean}(star feed)=81$kpc
kms$^{-1}$. This contrasts to the outflowing material, but also to gas
which is in the galaxy at this time and which remains in the galaxy as
gas at $z=0$, i.e. material which is neither being ejected nor feeding
star formation. Such ``retained'' gas has $jz_{mean}(retain)=67$kpc
kms$^{-1}$, with total angular momentum, $j_{mean}(retain)=127$kpc
kms$^{-1}$.

\subsection{What about mergers?}\label{bulge}

One of the characteristic features of CDM cosmologies is the merger
of galaxies. This process is ubiquitous, and essentially scale free
(e.g. \citealt{cole00,fakhouri10}), meaning that low mass field
galaxies which preferentially become discs dominated will have in
general similar merging histories as high mass galaxies, which have
bulges and may be dominated by spheroidal star systems.  Mergers are
widely expected to result in bulges.  In earlier simulations of galaxy
formation which suffered from overcooling, proto-galaxies rapidly
formed central baryonic concentrations through dynamical friction
\citep{navarro94} This problem was exacerbated in low resolution
simulations, which have denser central regions \citep{mayer08}.  As
explained above, feedback from supernovae prevents a dense bulge
forming in the central regions of the proto-galaxies in these current
simulations.  Further, as demonstrated in \cite{stewart09}, direct
accretion of stars is insignificant in low mass galaxies, whose
mergers are dominated by gas. The stars which are accreted in the
mergers, tend to get thrown into high energy orbits during the
interaction, meaning that the resulting spheroid is not centrally
concentrated. {\it A diffuse stellar halo rather than centrally
  concentrated bulge results from the scrambled trajectories of stars
  during the late major merger in our simulation}. Indeed, at z=0, our
simulated galaxy has an old (mean age 10.3 Gyrs), diffuse stellar
spheroidal component which comprises only $\sim 6\%$ of the stellar
mass of the simulated galaxy, even though it has no bulge. This is
consistent with observed low mass galaxies in all environments which
have old, dim spheroidal stellar populations in addition to their main
body of stars \citep{vansevicius04,hidalgo09,stinson09}.

Simulations, \citep{governato09,hopkins09} showed that, when mergers are gas rich and feedback is present,
bulges can be as little as 10\% of the resulting galaxy. In such gas
rich mergers, no strong bar forms to drive gas to the central
regions. Our simulated galaxy has a rich merger history including a
merger with mass ratio of 3:1 which has a final coalescence
$z\sim0.8$. All mergers are gas rich, with gas to stellar mass ratio
above one for all progenitors of our simulated galaxy. Several reasons
ensure that a bulge does not form during these mergers. Firstly, only
the inner cold gas is subject to being driven to the centre during the
mergers. The surrounding gas is not affected, but we know that inner
gas has preferentially low angular momentum. Secondly, in our
simulation as in \cite{hopkins09}, a bar does not form in gas
dominated mergers, limiting the amount of gas driven toward the
centre.

Thirdly, and most importantly, the feedback from the small fraction of
such gas which forms stars in a burst expels a large fraction of the
gas which has lost its angular momentum and has been driven toward the
centre of the galaxy, as reflected in the high mass loading during
this merging epoch, shown in Figure~\ref{massloading}.  This is clear
from the dip in the line traced by squares in Figure~\ref{jztime},
which indicates that the material which loses angular momentum
material is removed during the merger. Star formation is primarily
occurring in the very inner regions of the merging galaxies, and hence
it is coupled to the very material that has lost angular momentum
which is being blown out.  The effective coupling of SF with large
scale supernova driven outflows during merger induced star bursts is
well supported observationally \citep{martin99,hawthorn2007} and is
the extra ingredient in our models which allow us to form bulgeless
merger remnants.

\section{Discussion}\label{summary}

In our simulated dwarf galaxies, we are able to resolve the process of
energy injected from multiple overlapping supernovae, which occur in
star forming regions coupled to dense gas. The thermal energy injected
into the surrounding gas creates pressure, and is modelled to undergo
a blastwave phase which drives galactic winds. We emphasise that
neither degree nor direction of galactic winds are enforced {\it{a
    priori}} within our supernova feedback scheme.  It is
necessary for the spatial resolution to be significantly less than the
scale length of the disc to capture these effects \citep{colin10}.
More gas is expelled from our simulated galaxies than forms stars by
$z=0$,  up to 70\% of the original cosmic fraction.  Yet, similar to  observed isolated dwarfs \citep{geha06}, they remain gas dominated at all times, due to the continued cooling of gas into the disc and the low star formation rate. We have previously shown that our
model results in simulated dwarf galaxies which match multiple
properties of observed dwarf galaxies
(\citealt{governato10,oh10,brooks10}). Future and ongoing work will
include metal lines cooling at all temperatures, H$_2$ cooling and its
formation and destruction, and SF linked to the local H$_2$ fraction
\citep{booth07,robertson08,gnedin09}.  Preliminary results show that
the high density SF threshold adopted here closely mimics the
distribution of SF in simulations where SF is allowed only in gas with
a high fraction of H$_2$ (Christensen et al, in prep). The additional
cooling by metals and H$_2$ is counterbalanced by an increase in the
SN efficiency (Pontzen et al, in prep). None of these details of localised star formation affect the galaxy scale processes which are highlighted in this current study.   Here, we have outlined the
reasons that the galactic winds in our simulated galaxies are strongly
biased toward expelling low angular momentum gas. We identified the
interlinked processes responsible for such bias, which naturally occur
when feedback is effectively modelled in a hierarchically assembled
galaxy in a CDM Universe, namely:

\begin{itemize}
\item{\it Low angular momentum material is accreted and rapidly expelled early, when SF peaks}
\item{\it A low potential at early times favours early gas ejection.  }
\item{\it Later accreted material has higher angular momentum and forms a disc.}\\

  These ideas are similarly outlined by \cite{dekel86,binney01}, where a notion of expelling gas early is invoked to explain the angular momentum of disc galaxies.  Early ejection of material  is also postulated in recent models    of \cite{dutton09} to explain the mass distribution of disc galaxies.\\

\item{\it A reservoir of high angular momentum material exists beyond the star forming regions.}\\

Disc galaxies have extended HI discs \citep{broeils97}, as well as warm/hot gas in the halo \citep{spitzer56,spitzer75,sembach03}. These act as reservoirs of high angular momentum gas, which are beyond the star formation radius and thus not subject to being directly expelled from the galaxy.\\

\item{\it Outflows occur perpendicular to the disc, entraining relatively low angular momentum material and preventing infall of material from regions above and below the disc plane. }\\

Observations strongly indicate that galactic winds expel material perpendicular to galactic discs \citep{heckman87,shopbell98,veilleux02,hawthorn2007}, as predicted by theory \citep{maclow99,pieri07}. Gas which builds the disc at late times is accreted from regions that are relatively aligned to  the disc plane.\\

\item{\it Mergers or instabilities which  cause gas to lose angular momentum trigger star bursts which expel most of this low angular momentum gas. }\\

Star burst galaxies are generally associated with mergers, and  are observed to have significant outflows  \citep{heckman90,griffiths00,veilleux05} in local galaxies and at high redshift \citep{pettini2000,nestor10}, predominately driven from the inner starburst regions  \citep{cc85,strickland07}, ensuring that it is the very material that has lost  angular momentum during the merger which is being expelled. High velocity outflows from post-starburst galaxies are also observed at high redshift \citep{tremonti07}. The high mass loading in these events prevents large amounts of stars being formed from this material which had been driven toward the galaxy's inner regions. Feedback from central black holes/Active Galactic Nuclei (AGN) would also aid in expelling material which has lost angular momentum and has been driven to the central regions. There is evidence to suggest that such feedback is  directed perpendicular to the disc \citep{Irwin92,su10}, enhancing the processes outlined in this study.\\

\item{\it A diffuse stellar halo rather than a centrally concentrated bulge  results from the scrambled trajectories of stars during the late major merger in our simulation.}\\

The stellar rather than gas component of mergers between progenitors which are massive enough to contain stars themselves, but which do not themselves have bulges, result in a diffuse halo component. 
\end{itemize}

By expelling large amounts of preferentially low angular momentum gas
from the galaxy, particularly at early times and during merger events,
the shape of the angular momentum distribution of baryons is altered
significantly from that of their parent dark matter halos. This allows
the formation of bulgeless disc galaxies. Future work is needed to see
if the same processes (possibly in conjunction with black hole/AGN  related feedback which also emanates from central regions and thus expels low angular momentum gas)  is able to
explain the suppression of bulges  in more massive halos and
the formation of bright disc galaxies with small classical bulges
\citep{kormendy10,peebles10}.

\section*{Acknowledgments}
CB and BKG acknowledge the support of the UKs Science \&
Technology Facilities Council (STFC Grant ST/F002432/1).
CB and GS thank the DEISA consortium, co-funded
through EU FP6 project RI-031513 and the FP7 project
RI-222919 for support within the DEISA Extreme Comput-
ing InitativeFG, and TQ were supported by NSF ITR grant
PHY- 0205413. FG acknowledges support from a Theodore
Dunham grant, HST GO-1125, NSF grant AST-0607819 and
NASA ATP NNX08AG84G. AB acknowledges support from
the Sherman Fairchild Foundation. Simulations were run at
TACC, ARSC, and NAS. We acknowledge the computa-
tional support provided by the UKÕs National Cosmology
Supercomputer, COSMOS.

\bibliographystyle{mn2e}
\bibliography{brookbib}

\end{document}